\documentclass[11pt,twoside]{article}
\usepackage{asp2010}

\resetcounters
\aspvolume{455}
\aspcpryear{2012}
\aspvoltitle{4$^{th}$ {\em Hinode} Science Meeting: Unsolved 
Problems and Recent Insights}
\aspvolauthor{L.~R.~Bellot Rubio, F.~Reale, and M.~Carlsson, eds.}

\begin{document}

\title{Prominence Seismology}
\author{I.~Arregui,$^1$ J.~L.~Ballester,$^1$ R.~Oliver,$^1$ R.~Soler,$^2$ and J.~Terradas$^1$
\affil{$^1$Departament de F\'{\i}sica, Universitat de les Illes Balears, 07122 Palma de Mallorca, Spain}
\affil{$^2$Centrum voor Plasma Astrofysica, K.U.~Leuven, Celestijnenlaan 200B,  B-3001 Leuven, Belgium}}

\begin{abstract}
Given the difficulty in directly determining prominence physical
parameters from observations, prominence seismology stands as an
alternative method to probe the nature of these structures.  We show
recent examples of the application of magnetohydrodynamic (MHD)
seismology techniques to infer physical parameters in prominence
plasmas.  They are based on the application of inversion techniques
using observed periods, damping times, and plasma flow speeds of
prominence thread oscillations. The contribution of {\em Hinode} to
the subject has been of central importance.  We show an example based
on data obtained with {\em Hinode}'s Solar Optical
Telescope. Observations show an active region limb prominence,
composed by a myriad of thin horizontal threads that flow following a
path parallel to the photosphere and display synchronous vertical
oscillations. The coexistence of waves and flows can be firmly
established. By making use of an interpretation based on transverse
MHD kink oscillations, a seismological analysis of this event is
performed.  It is shown that the combination of high quality {\em Hinode}
observations and proper theoretical models allows flows and waves to
become two useful characteristics for our understanding of the nature
of solar prominences.
\end{abstract}

\section{Introduction}

Solar prominences are one the most intriguing manifestations of solar
activity. These structures consist of large clouds of plasma, two
orders of magnitude cooler and denser than the surrounding corona,
suspended against gravity by forces thought to be of magnetic origin.
The physical properties of prominence plasmas are akin to those of the
chromospheric plasma, hence some as yet not well determined mechanisms
must provide the required thermal isolation and hydrodynamic support
during lifetimes that last from days to weeks.  The nature of solar
prominences is closely linked to their sub-resolution structuring.
Early studies by \cite{1959HDP....52...80D} and
\cite{1967SoPh....2...39K} already pointed out that prominences are
composed by many fine threads. This has been confirmed by more recent
high-resolution observations obtained by, e.g.,
\cite{2005SoPh..226..239L}. The fine threads are made of cool
absorbing material, believed to outline magnetic flux tubes
\citep{2008SoPh..250...31M}. Their average width is about 0\farcs4,
their length is in between 5\arcsec\/ and 40\arcsec\/, and they have
lifetimes of up to 20 minutes.

The measurement of physical properties of prominence plasmas by direct observation is challenging. 
An alternative to gain knowledge about the physical conditions in prominences is the combined use of observed and theoretical 
wave properties. The presence of waves and oscillations in solar prominences is known since a long time \citep[see reviews by][]{2009SSRv..149..175O,2010AdSpR..46..364B}. Individual threads or groups of them oscillate with  periods that range between 3 and 20 minutes \citep{1991SoPh..134..275Y,1991SoPh..132...63Y, 2007SoPh..246...65L,2007Sci...318.1577O, 2008ApJ...678L.153T,2009A&A...499..595N,2009ApJ...704..870L} . A recurrently observed property of prominence oscillations is their rapid temporal damping, with perturbations decaying in time-scales of only a few oscillatory periods \citep[see][for recent examples]{2002A&A...393..637T,Linthesis,2009A&A...499..595N}. Another relevant characteristic of waves  in solar prominences is the ubiquitous 
presence of flows \citep{1998Natur.396..440Z,2003SoPh..216..109L,2005SoPh..226..239L}. Flow speeds in the range 5--25 km~s$^{-1}$ are usually reported in quiescent filament threads, while in active region threads flow speeds of up to 50 km~s$^{-1}$ are detected \citep{2007Sci...318.1577O}.

\begin{figure}[t]
\centering
  \includegraphics[width=11.5cm]{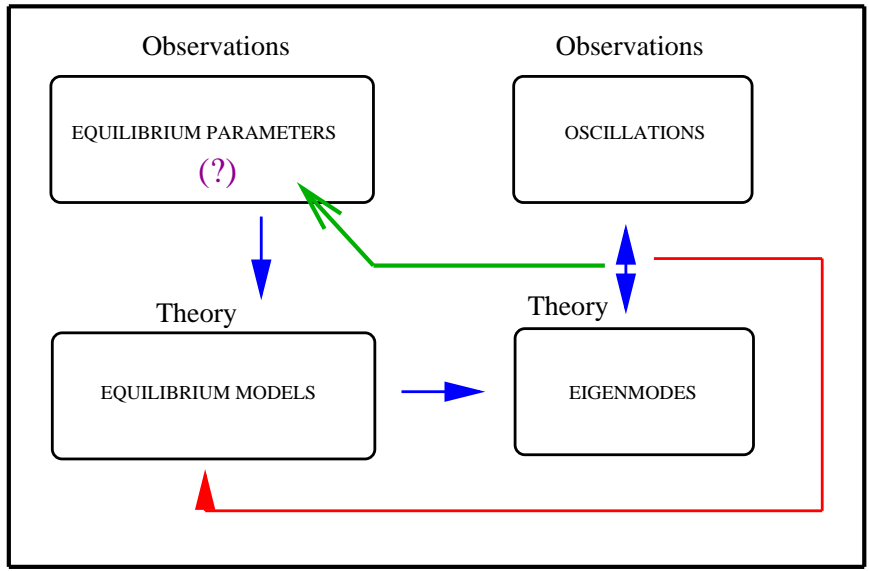} 
   \caption{Systematic of MHD seismology.}
   \label{fig:sismology}
\end{figure}

Transverse thread oscillations are commonly interpreted in terms of
standing or propagating MHD kink waves. The MHD wave interpretation of
thread oscillations has allowed the development of theoretical models
\citep[see reviews by][]{2005SSRv..121..105B,2006RSPTA.364..405B}.
Theoretical damping mechanisms have also been developed
\citep[see][for recent
reviews]{2010AdSpR..46..364B,2011SSRv..158..169A}. As information on
geometrical properties, damping mechanisms, and flows are incorporated
in theoretical models for MHD wave studies, more refined results are
obtained that allow a better implementation of prominence seismology.

Solar atmospheric seismology aims to determine physical parameters in
magnetic and plasma structures that are difficult to measure by direct
means. It is a remote diagnostics method that combines observations of
oscillations and waves in magnetic structures, together with
theoretical results from the analysis of oscillatory properties of
models of those structures.  It was first suggested by
\cite{1970PASJ...22..341U} and \cite{1984ApJ...279..857R}, in the
coronal context, and by \cite {1995ASSL..199.....T} in the prominence
context.  The systematic of MHD seismology is displayed in
Figure~\ref{fig:sismology}. Observations of solar coronal magnetic and
plasma structures provide us with information from which theoretical
models about their equilibrium can be constructed. On the other hand,
observations also provide us with measurements of certain properties,
such as periods, damping times, or flow speeds. By analyzing the wave
properties of given theoretical models they can be compared to the
observed wave properties. If we find a perfect agreement between
observed and theoretical wave properties we can aim to derive some
unknown physical parameters of interest. Meanwhile, we can test and
improve our models or find constraints. This article discusses three
tools for the application of MHD prominence seismology, with an
emphasis on the {\em Hinode} contribution to the area.

\section{Seismology Using Observed Periods of Thread Oscillations}

A recent application of the prominence seismology technique, using the
period of observed filament thread transverse oscillations can be
found in \citet{2009ApJ...704..870L}. These authors find observational
evidence about swaying motions of individual filament threads from
high resolution observations obtained with the Swedish 1-m Solar
Telescope in La Palma. The presence of waves propagating along
individual threads was already evident in, e.g.,
\citet{2007SoPh..246...65L}. However, the fact that line-of-sight
oscillations are observed in prominences beyond the limb, as well as
in filaments against the disk, suggests that the planes of the
oscillation may acquire various orientations relative to the local
solar reference system. For this reason, \citet{2009ApJ...704..870L}
combine simultaneous recordings of motions in the line of sight and in
the plane of the sky, which leads to information about the orientation
of the oscillatory plane in each case. Periodic oscillatory signals
are obtained in a number of threads, that are next fitted to sine
curves, from which the period and the amplitude of the waves are
derived. The presence of different cuts along the structures allow
\cite{2009ApJ...704..870L} to obtain the phase difference between the
fitted curves, which can be used to measure the phase velocities of
the waves. The overall periods and mean velocity amplitudes that are
obtained correspond to short period, $P\sim$ 3.6 minutes, and small
amplitude $\sim 2$ km~s$^{-1}$ oscillations. The information obtained
from these H$\alpha$ filtergrams in the plane of the sky is combined
with H$\alpha$ Dopplergrams, which allow to detect oscillations in
the line-of-sight direction.  By combining the observed oscillations
in the two orthogonal directions the full vectors are derived, which
show that the oscillatory planes are close to the vertical.

\cite{2009ApJ...704..870L} interpret the observed swaying thread 
oscillations as MHD kink waves supported by the thread body. By
assuming the classic one-dimensional, straight, flux tube model a
comparison between the observed wave properties and the theoretical
prediction can be made in order to obtain the physical parameters of
interest, namely the Alfv\'en speed and the magnetic field strength in
the studied threads. To this end the observed phase speed is directly
associated to the kink speed, which in the limit of high density
contrast, typical of filament plasmas, is simply reduced to
$c_k\simeq\sqrt{2}v_{Ai}$, with $v_{Ai}$ the prominence Alfv\'en speed.
This allows to obtain the thread Alfv\'en speed through $v_{Ai}\simeq
V_{\rm phase}/\sqrt{2}$. The obtained values for a set of 10 threads
can be found in Table~2 in \cite{2009ApJ...704..870L}. Once the
Alfv\'en speed in each thread is determined, the magnetic field
strength can be computed, if a given internal density is assumed. For
a typical value of $\rho_i=5\times10^{-11}$ kg m$^{-3}$ magnetic field
strengths in between 0.9--3.5~G are obtained, for the analyzed
events. The important conclusion that we extract from the analysis by
\cite{2009ApJ...704..870L} is that prominence seismology is possible
and works well, provided high resolution observations are available.

\section{Seismology Using Observed Periods and Damping Times}

\begin{figure*}[t]
\centering
\hspace*{-1em}
  \includegraphics[height=5.5cm,width=6.5cm]{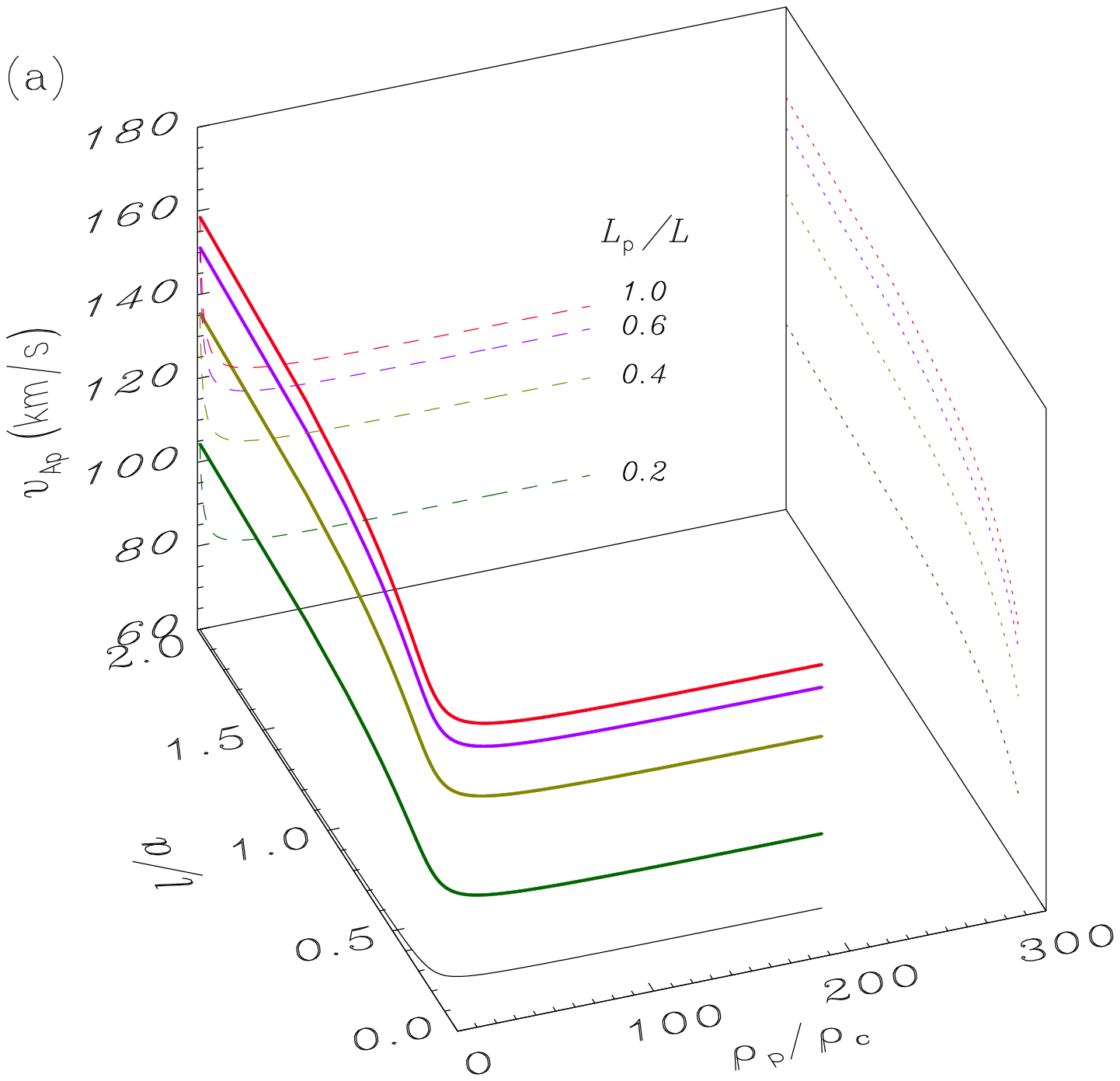}
  \includegraphics[height=5.5cm,width=6.5cm]{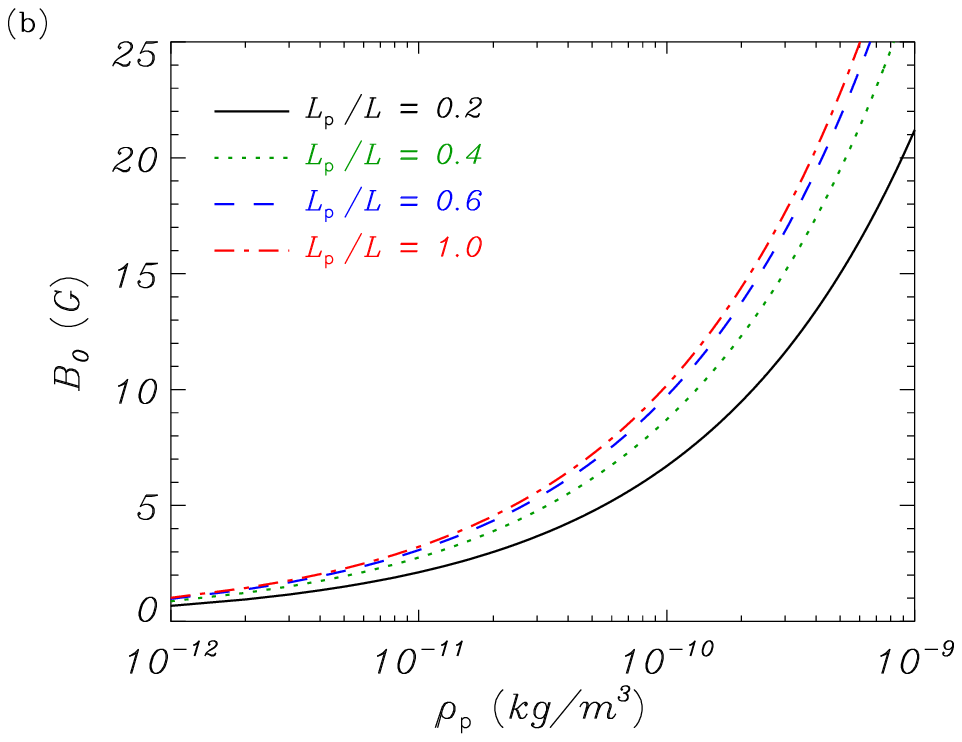}
  \caption{Determination of prominence Alfv\'en speed {\em (a)} and
  magnetic field strength {\em (b)} from the computation of periods
  and damping times for standing kink oscillations in two-dimensional
  prominence thread models and observations of period and damping
  times in transverse thread oscillations. The observed period and
  damping time are 20 and 60 minutes, respectively, and $L=10^5$~km.}
  \label{fig:seismology}
\end{figure*}

The damping of prominence oscillations is a clear feature in many
observed events. \cite{2009ApJ...704..870L} show that the amplitudes
of the waves passing through two different cuts along a thread are
notably different.  Among the different damping mechanisms that have
been put forward in order to explain the damping of MHD oscillations
in prominence plasmas resonant absorption in the Alfv\'en continuum
seems a very plausible one. The mechanism relies on the non-uniformity
of the medium in the transverse direction. It was suggested to explain
the damping of transverse kink waves in prominence threads by
\cite{2008ApJ...682L.141A}. The damping of wave modes can be used as
an additional source of information about the physical properties of
prominence plasmas. In the context of transversely oscillating coronal
loops this was done by \cite{2007A&A...463..333A} and
\cite{2008A&A...484..851G}.  Their analytical and numerical inversion
schemes make use of the simple idea that it is the same magnetic
structure, whose equilibrium conditions we are interested to assess,
that is oscillating with a given period and undergoing a given damping
rate.  By computing the kink normal mode frequency and damping time as
a function of the relevant equilibrium parameters for a
one-dimensional model, the period, $P$ and damping ratio, $P/\tau_d$,
have the following dependencies
\begin{eqnarray}\label{relations}
P=P(k_z,c,l/a,v_{Ai}), \mbox{\hspace{1cm}} \frac{P}{\tau_d}=\frac{P}{\tau_d}(k_z,c,l/a),
\end{eqnarray}
with $v_{Ai}$ the internal Alfv\'en speed, $k_z$ the longitudinal
wavenumber, $c=\rho_i/\rho_e$ the density contrast, and $l/a$ the
transverse inhomogeneity length-scale, in units of the tube radius,
$a$. In the case of coronal loop oscillations, an estimate for $k_z$
can be obtained directly from the length of the loop and the fact that
the fundamental kink mode wavelength is twice this quantity. For
filament threads, the wavelength of oscillations needs to be
measured. Relations~\ref{relations} indicate that, if no assumption is
made on any of the physical parameters of interest, we have two
observed quantities, period and damping time, and three unknowns,
density contrast, transverse inhomogeneity length-scale, and Alfv\'en
speed. There are therefore infinite different equilibrium models that
can equally well explain the observations.  These valid equilibrium
models delineate a one-dimensional solution curve in the
three-dimensional parameter space ($c$, $l/a$, $v_{Ai}$). When
partially filled threads, with the dense part occupying a length $L_p$
shorter than the total length of the tube $ L$ are considered, the
period and damping time of thread oscillations are seen to depend on
$L_p/L$ \citep{2010ApJ...722.1778S,2011A&A...533A..60A}. Then, one of
such curves is obtained for each value of the length of the thread.
The solutions to the inverse problem are shown in
Fig.~\ref{fig:seismology}a for a set of values for $L_p$. It can be
appreciated that, even if each curve gives an infinite number of
solutions, they define a rather constrained range of values for the
thread Alfv\'en speed. This figure also shows that the ratio $L_p/L$
is a fundamental parameter in order to perform an accurate seismology
of prominence threads. Because of the insensitiveness of the damping
ratio with density contrast, for the typically large values of this
parameter in prominence plasmas, the obtained solution curves display
an asymptotic behavior for large values of $c$. This allows us to
obtain precise estimates for the filament thread Alfv\'en speed and
the transverse inhomogeneity length scale in that limit for each of
the curves. The computation of the magnetic field strength from the
obtained seismological curve requires the assumption of a particular
value for either the filament or the coronal density. The resulting
curves for a typical coronal density and several values of $L_p/L$ are
shown in Figure~\ref{fig:seismology}b. As can be seen, precise values
of the magnetic field strength cannot be obtained, unless the density
contrast is accurately known.

\section{Seismology of Flowing and Oscillating Prominence Threads}

The question of whether propagating disturbances detected in
chromospheric and coronal structures are waves or flows has become a
hot topic that demands for both observational and theoretical
assessment in order to perform an accurate seismology. The
contribution of {\em Hinode} to the subject has been of central importance.
Flows constitute and additional source of information about the
physical conditions of oscillating structures in conjunction to
oscillating periods and damping times.  The first seismological
application of prominence seismology using {\em Hinode} observations of
flowing and transversely oscillating threads was presented by
\cite{2008ApJ...678L.153T}, using observations obtained in an active
region filament by \cite{2007Sci...318.1577O}. The observations show a
number of threads that flow following a path parallel to the
photosphere while they are oscillating in the vertical direction. The
relevance of this particular event is in the fact that the coexistence
of waves and flows can be firmly established, so there is no ambiguity
about the wave or flow character of a given dynamic feature. We have
both of them in this particular example. \cite{2007Sci...318.1577O}
analyze 6 threads and Table~1 displays the relevant measured
quantities.

A detailed view of the event analyzed by \cite{2007Sci...318.1577O}
shows that when a given thread is selected and several cuts along its
structure are analyzed as a function of time, oscillations that are
synchronous along the entire length of the thread are found. This
means that the maximum and minimum amplitudes occur at nearly the same
time for all locations. This led \cite{2008ApJ...678L.153T} to
interpret these oscillations in terms of the kink mode of a magnetic
flux tube.

% Requires the booktabs if the memoir class is not being used
\begin{table}[ht]
\tabcolsep .68em
   \centering
   \caption{Summary of geometric and wave properties of vertically oscillating flowing threads analyzed by \citet{2007Sci...318.1577O}. $2W$ is the thread length, $v_0$ its horizontal flow velocity, $P$ the oscillatory period, $V$ the oscillatory velocity amplitude, and $H$ the height above the photosphere.\label{table1}}
   \label{tab:booktabs}
   \begin{tabular}{cccccc}% Column formatting, @{} suppresses leading/trailing space
\noalign{\smallskip}
\hline 
\noalign{\smallskip}
Thread & $2W$ (km) & $v_0$ (km~s$^{-1}$) & $P$ (s) & $V$ (km~s$^{-1}$) & $H$ (km) \\
\noalign{\smallskip}
\hline
\noalign{\smallskip}
1 & 3600 & 39 & 174 $\pm$ 25 & 16 & 18\,300 \\
2 & 16\,000 & 15 & 240 $\pm$ 30 & 15 & 12\,400 \\
3 & 6700 & 39 & 230 $\pm$ 87 & 12 & 14\,700 \\
4 & 2200 & 46 & 180 $\pm$ 137 & 8 & 19\,000 \\
5 & 3500 & 45 & 135 $\pm$ 21 & 9 & 14\,300 \\
6 & 1700 & 25 & 250 $\pm$ 17 & 22 & 17\,200 \\
\noalign{\smallskip}
\hline
       \end{tabular}
\end{table}

\begin{figure*}[t]
\centering
\hspace*{-1.5em}
  \includegraphics[height=12cm,width=11.7cm]{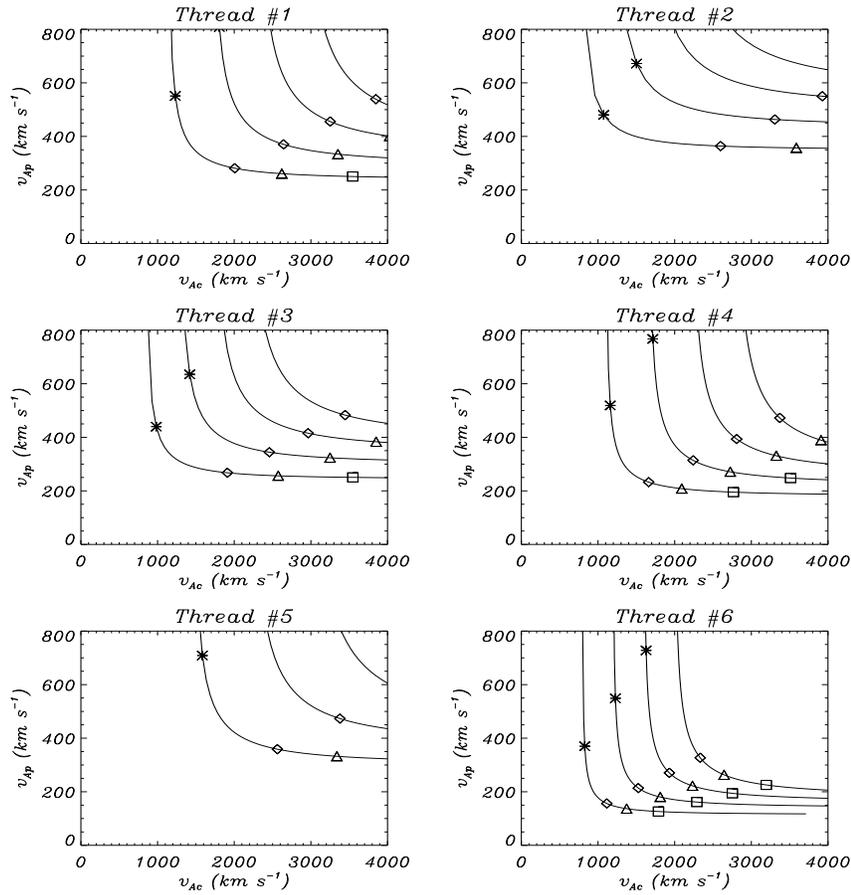}
  \caption{Dependence of the Alfv\'en velocity in the thread as a
  function of the coronal Alfv\'en velocity for the six threads
  observed by \citet{2007Sci...318.1577O}. In each panel, from bottom
  to top, the curves correspond to a length of magnetic field lines of
  100\,000, 150\,000, 200\,000, and 250\,000~km,
  respectively. Asterisks, diamonds, triangles, and squares correspond
  to density ratios of the thread to the coronal gas
  $\rho_p/\rho_c\simeq 5,50,100, 200$.}  \label{fig:hinode}
\end{figure*}

Let us first neglect the presence of mass flows. By using previous
theoretical results from a normal mode analysis in a two-dimensional
piecewise filament thread model by \cite{2002ApJ...580..550D} and
\cite{2005SoPh..229...79D}, \cite{2008ApJ...678L.153T} find that,
although it is not possible to univocally determine the physical
parameters of interest, a one-to-one relation between the thread
Alfv\'en speed and the coronal Alfv\'en speed can be established. This
relation comes in the form of a number of curves relating the two
Alfv\'en speeds for different values of the length of the magnetic
flux tube and the density contrast between the filament and coronal
plasma.  Figure~\ref{fig:hinode} shows the derived values by changing
the length of the tube from bottom to top and the density contrast,
from left to right. An interesting property of the obtained curves is
that they display an asymptotic behavior for large values of the
density contrast, typical of filament to coronal plasmas, and hence a
lower limit for the thread Alfv\'en speed can be obtained. Take for
instance thread \#6.  Considering a length of the total magnetic flux
tube of $L=100$ Mm, an overall value between 120 and 350~km~s$^{-1}$
for the thread Alfv\'en speed is obtained.

Next mass flows are considered. First a simple approximation is
made. Consider the form in which mass flow along the cylinder, $v_0$,
enters in the linear MHD waves equations through the differential
operator
\begin{displaymath}
\frac{\partial}{\partial t}+ v_0\frac{\partial}{\partial z}.
\end{displaymath}
The terms coming from the equilibrium flow can, in a first
approximation, be ignored because, as noted by
\cite{2005SoPh..229...79D}, inside the cylinder the terms with
derivatives along the tube are much smaller than those with radial or
azimuthal derivatives. By following this approach the problem reduces
to solving a time-dependent problem with a varying density profile,
$\rho(z,t)$, representing a dense part moving along the tube with the
flow speed. After solving the two-dimensional wave equations one finds
that the flow velocities measured by \citet{2007Sci...318.1577O}
result in slightly shorter kink mode periods than the ones derived in
the absence of flow. Differences are however small and we find period
shifts in between 3 and 5\%.

Finally, a more complete approach to the problem has been followed by
\citet{2008ApJ...678L.153T} who consider the numerical solution of the
full MHD wave equations, with no further approximations. The thin tube
approximation is not used, the flow is maintained in the
equations. Also the density is allowed to change in the simulation as
a nonlinear code is used. The full numerical result confirms the
previous approximate results regarding the effect of the flow on the
obtained periods, and therefore, on the derived Alfv\'en speed values.

\section{Summary}

This paper presents recent developments of the MHD seismology
technique applied to prominence plasmas.  The combination of refined
theoretical models for prominence thread oscillations together with
observations allows for the estimation of difficult to measure
quantities in these objects. Three complementary techniques for the
inversion of physical parameters have been discussed. They make use of
the MHD kink wave interpretation for transverse oscillations in
filament and active region threads together with observational
estimates for quantities such as periods, damping rates, and mass flow
speeds. In general, the solution to the inverse problem is unable to
provide a single value for all the parameters of interest, since the
quantity of unknowns outnumbers that of measured wave
properties. However, estimates of physical parameters in a narrow
range of values can be derived.  A proper further development of
prominence seismology requires improvements in both theory and
observations. Our example using {\em Hinode} observations demonstrates that
high quality observations and proper theoretical analysis allow flows
and waves to become two useful characteristics for our understanding
of the nature of solar prominences.

\acknowledgements The authors
acknowledge the funding provided under the pro\-ject AYA2006-07637 by
Spanish MICINN and FEDER Funds. RS acknowledges a postdoctoral
fellowship within the EU research and training network SOLAIRE.

%\bibliography{hinode4} 
\end{document}